\documentclass[useAMS,usenatbib]{mn2e}

\usepackage{graphicx}
\usepackage{natbib}

\title[Clumpy Disc and Bulge Formation]{Clumpy Disc and Bulge Formation}
\author[Perez et al.]{Josefa Perez$^{1,2,3,}$, Octavio Valenzuela$^{4}$, Patricia B. Tissera$^{1,3}$ and Leo Michel-Dansac$^{5,6}$\\
$^{1}$Instituto de Astronom\'\i a y F\'\i sica del Espacio,Conicet-UBA, CC67, Suc.28,Ciudad de  Buenos Aires, Argentina.\\
$^{2}$C\'atedra de Astronom\'\i a Estelar.  Facultad de Ciencias Astron\'omicas y Geof\'\i sicas. Universidad Nacional de La Plata, Argentina.\\
$^{3}$Consejo Nacional de Investigaciones Cient\'\i ficas y T\'ecnicas, CONICET, Argentina.\\
$^{4}$Instituto de Astronom\'ia, Universidad Nacional Aut\'onoma de M\'exico, A.P. 70-264,
   04510, M\'exico, D.F.; Universitaria, D.F., M\'exico.\\
$^{5}$Centre de Recherche Astrophysique de Lyon, Universit\'e de Lyon,
  Universit\'e Lyon 1, Observatoire de Lyon, France.\\
$^{6}$ Ecole Normale Sup\'erieure de
  Lyon, CNRS, UMR 5574, 9 avenue Charles Andr\'e, Saint-Genis Laval, 69230,
  France.\\
}

\begin{document}

\date{Accepted ????. Received ????; in original form ????}
\pagerange{\pageref{firstpage}--\pageref{lastpage}} \pubyear{2002}
\maketitle 

\label{firstpage}

\begin{abstract}
We present a set of hydrodynamical/Nbody controlled simulations 
 of isolated gas rich galaxies that self-consistently include SN feedback and 
a detailed chemical evolution model, both tested
in cosmological simulations. The initial conditions are motivated 
by the observed star forming galaxies at z $\sim$ 2-3. 
We find that  the presence of a multiphase interstellar media in our models 
promotes the growth of disc instability favouring the formation
of clumps which in general, are not easily disrupted 
on timescales compared to the migration time. 
We show that stellar clumps  migrate towards 
the central region and contribute to form a classical-like bulge with a
Sersic index,  n $> 2$. 
 Our physically-motivated Supernova feedback  has a mild influence on clump survival and evolution, 
 partially limiting the mass growth of clumps as the energy released per Supernova 
event is increased, with the consequent flattening of the bulge profile. 
 This  regulation does not prevent the building of a classical-like bulge 
even for the most energetic feedback tested. 
Our  Supernova feedback model is able
to establish a self-regulated star formation, producing mass-loaded outflows  
and stellar age spreads comparable to observations. 
We find that the bulge formation by clumps may coexit with other channels of bulge assembly such as bar and mergers.
Our results suggest that  galactic bulges could be interpreted as composite systems  
with structural components and  stellar populations 
 storing  archaeological information of the  dynamical history of their galaxy.
\end{abstract}


\section{Introduction}

The formation of bulges is still a matter of large debate. 
 Observations suggest that the Sersic index ({\it n}) of a large sample 
of bulges are smaller than what previously thought, 
 specifically for late type galaxies, suggesting  
that secular processes play an important role in formation of this component. 
These observations  establish a clear dichotomy in bulge properties, 
classifying them into {\it classical-} ({\it n} $>$ 2) 
and {\it pseudo-}  ({\it n} $<$ 2) {\it bulges} 
 \citep{fisher08}.
Mergers are considered as one possible mechanism of formation as well as secular evolution. 
Both formation modes seem to generate bulges with different properties 
 in numerical simulations, being the first ones more prompted to yield 
de Vaucouleurs (high-{\it n} ) density profiles compared to the second ones 
which would favour more exponential ones (low-{\it n}).  
  However,  there is currently not theory to explain 
the Sersic index values in bulges.   Recently, a  third possibility 
has came out from new observations of high redshift galaxies and high resolution, improved numerical models.

The  high-redshift observations show that 
the most intense star formers in the Universe are galaxy discs 
at z$\sim$2. 
Their morphologies are consistent with 
thick, gas-rich discs, rotationally supported with 
circular velocities of $\sim 200$ km/s.   
 A peculiar feature of these galaxies is that their discs 
show the presence of several giant clumps of $\sim 10^{8}- 10^{10} M_{\odot}$ 
\citep[e.g.][]{Forster2011} where an intense star formation is observed  
  \citep[e.g.][]{forster09}. 
Gravitational instabilities in gas-rich turbulent discs have been proposed 
as the fundamental mechanism to account for this fragmentation. 
This scenario is also able to explain the bulge formation 
via clump migration  \citep{elmegreen08, genzel08, bournaud11, guo11}. 

Cosmological simulations \citep{ceverino10}, 
analytical works \citep{dekel09} and simulated idealized discs 
in isolation \citep{immeli04, bournaud07, bournaud11} 
have succeeded to explain most of the observed clump properties. 
 However, a remaining aspect should be addressed, that is, the ability of clumps to 
survive  the effect of SN feedback. 
Several works have disregarded  clump disruption by 
SN thermal heating, claiming  radiation pressure to play  a more dominant role  
\citep{dekel09, murray10}.  
Hydrodynamical simulations including radiation pressure models have found that clumps disrupt   
before coalescing the bulge \citep{hopkins11, genel12}. 
 However,  \citet{krumholz10} use analytical models to conclude 
that radiation pressure would not be efficient to disrupt clumps in high redshift galaxies. 
The strongest evidences  suggesting that clumps survive long enough 
 to reach the bulge are  the observational estimations of clump ages \citep{genzel11}, 
and the   radial gradients detected in  the clump properties \citep{guo11}.

In this paper, we revise the clump survival problem within the context 
of our adopted SN feedback and multiphase model, and analyse the 
impact that clumps might have on bulge formation. 
For this purpose, we use hydrodynamical simulations of isolated gas-rich discs and 
a realistic physically-motivated SN feedback model implemented with a multiphase treatment 
of the ISM \citep{scannapieco05, scannapieco06}. 
We also  explore how different mechanisms 
of bulge formation (clumps migration, mergers, bars) 
 compete with each other, and trace their structural and stellar population properties. 
These last results could help to unravel the dynamical galactic history 
contained in the archaeological information of bulges.

\section{Numerical simulations and clump identification}\label{s1}

We analysed a set of simulations  run by using an extended version of the {\small GADGET-2} code 
 which includes 
a realistic SN feedback model, implemented with 
a multiphase treatment of the ISM \citep{scannapieco05, scannapieco06}. This code 
allows the coexistence of gas clouds with different 
thermodynamical properties and is able to describe 
the injection of energy into the ISM producing the 
self-regulation of the SF  and  the triggering of mass-loaded galactic outflows, 
without the need to introduce mass-dependent parameters, or 
to change discontinuously particle momentum to start a wind  
or  temporary suppression of the radiative cooling.  The radiative cooling rates
are estimated according to the metallicity of the gas.

We studied a set of hydrodynamical simulations  of 
pre-prepared disc galaxies initially composed  of a dark matter halo (following 
a NFW profile), a Hernquist bulge  and 
 an exponential disc, with a total baryonic mass of $M_{\rm b} \sim 5 \times 
 10^{10} M_{\odot}$. 
All experiments  were run with a 50 percent of discs in form of gas in order 
to reproduce observations of z$~\sim~$2 gas-rich 
galaxies (Daddi et al. 2010). 
We use $200\,000$ dark matter particles, $100\,000$ stars initially distributed in the 
stellar disc and bulge, and $100\,000$ initial gas particles 
in the disc component with a mass resolution  of $\approx 2 \times 10^5$ M$\odot$. 
A gravitational softening of $\epsilon_{\rm G}=0.16$ kpc was adopted for 
the gas particles, $\epsilon_{\rm S}=0.20$ kpc for the stars, and 
$\epsilon_{\rm DM}=0.32$ kpc for the dark matter. 
The initial metallicity of the gas has been set so that the  simulated discs lie on the mass-metallicity relation at $z \sim 2$
as explained in \citet{perez11}.

We analysed five simulations of the  isolated gas-rich disc  
 varying the ISM and the SN feedback models.
Three simulations, S.FeMu, S.ModFeMu and S.StrFeMu, have 
the same multiphase ISM model but explore   
different SN feedback efficiencies, parametrized by 
their energy release, E$_{SN}$. 
The other two simulations, S.Mu and S.BasicSPH, do not include SN feedback. The former 
includes the  multiphase ISM model of \citet{scannapieco05} but without SN 
feedback, and the latter has  star formation but 
with no SN feedback nor multiphase ISM. 
We also study the clump growth  during galaxy interaction and bar formation using   
two simulations: S.FeMu$_{-}$Int and S.FeMu$_{-}$Bar, respectively. 
 We note that the interacting case and its associated isolated simulation  
(S.FeMu$_{-}$Int and S.FeMu) were previously used 
by \citet{perez11} to investigate the evolution of metallicity gradients 
in high-z galaxies\footnote{This simulation was referred as SimVI by \citet{perez11}.}.  
S.FeMu$_{-}$Bar is a version of S.FeMu with a shorter 
radial scale length in order to have a centrally dominating disc prone to bar instability.  
The main parameters of the analysed simulations are summarized in Table~\ref{tab1} for the initial conditions and
the after $\approx 3.5$ Gyr of evolution. 

Consistently with previous works, our simulated discs 
fragment into large clumps. In order to identify them, we use 
a morphological criteria similar to that used by \citet{bournaud07},  
based on the fact that clumps represent local over-densities. 
First, we  compute the face-on projected 
surface density on a polar grid, defined to control the particle noise. 
Then, we keep the pixels which represent over-densities 
compared to the average density at the same radius. 
This over-density criteria is controlled by a free parameter, 
set to eliminate extended connected regions as spiral arms. 
Selected clumps have masses ranging from $\approx 10^7 M\odot$ 
to $ \approx 10^{9} M\odot$ with a mean at  
$\sim 10^8$ M$\odot$ (see Fig.~\ref{models2}), 
which are in agreement with previous observational and numerical works. 

\begin{table*}
  \begin{center}
    \caption{Main parameters of the numerical experiments. 
     E$_{SN}$ is amount of SN energy release by each event in units of $10^{51}$ erg s$^{-1}$. 
  {\it   n}$_{\rm initial}$ and {\it n}$_{\rm final}$  
are  the initial and final Sersic indexes of the bulges, and  
{\it   Re}$_{\rm \, initial}$ and {\it Re}$_{\rm \, final}$ the respective effective radius.  (B/T) is 
the final bulge-to-total stellar masses, where the bulge is computed within 2Re. Errors correspond to one standard deviation.}
    \label{tab1}
    \begin{tabular}{|l|c|c|c|c|c|c}\hline
       {Simulations} & E$_{SN}$  & {\it n}$_{\rm initial}$ & {\it n}$_{\rm final}$ & {\it Re}$_{\rm \, initial}$ & {\it Re}$_{\rm \, final}$   & B/T\\
      \hline
       S.BasicSPH   & $-$                      & 1.4 $\pm$ 0.2 & 1.6 $\pm$ 0.2   & 0.63 $\pm$ 0.01 &  0.67 $\pm$ 0.02   & 0.24\\
       S.Mu        & $-$                      & 1.3 $\pm$ 0.2 & 3.5 $\pm$ 0.1   & 0.61 $\pm$ 0.02 &  0.33 $\pm$ 0.01   & 0.32\\
       S.FeMu      & $0.5$                    & 1.4 $\pm$ 0.1 & 4.2 $\pm$ 0.1   & 0.65 $\pm$ 0.02 &  0.33 $\pm$ 0.02   & 0.31\\
       S.ModFeMu   & $0.7$                    & 1.3 $\pm$ 0.2 & 3.2 $\pm$ 0.1   & 0.61 $\pm$ 0.02 &  0.40 $\pm$ 0.02   & 0.34\\
       S.StrFeMu   & $1.0$                    & 1.3 $\pm$ 0.2 & 2.9 $\pm$ 0.1   & 0.61 $\pm$ 0.02 &  0.58 $\pm$ 0.01   & 0.29\\
      \hline
       S.FeMu$_{-}$Int             & $0.5$                 & 1.3 $\pm$ 0.1 & 2.9 $\pm$ 0.3  & 0.53 $\pm$ 0.02 &  0.51 $\pm$ 0.02   & 0.39\\
       S.FeMu$_{-}$Bar & $0.5$       & 1.6 $\pm$ 0.2  & 2.3 $\pm$ 0.2     & 0.47 $\pm$ 0.02 &  0.30 $\pm$ 0.02    & 0.35\\
      \hline
    \end{tabular} 
  \end{center}
  \vspace{1mm}
\end{table*}

\section{Clumps Survival}

Our simulated gas-rich discs fragment into large clumps, where an 
intense star formation activity is detected. 
 In good agreement with observations \citep{guo11}, 
 we  find that the individual contribution of clumps 
to the global SFR of the host galaxy is in general lower than 10$\%$, 
with a collective contribution of $\approx 45 \%$ on average,  and a maximum of  $\sim 60\%$. 
Even with an active star formation in clumps, the rate at which 
they form stars is less than one percent of the clumps mass 
per free-fall time,  which determines a dimensionless  
star-formation rate efficiency, $e_{ff}$, 
to be lower than 0.01  \citep[see eq. 6 of][]{krumholz10}. 

\begin{figure*}
\centering
\includegraphics[width=75mm,height=55mm]{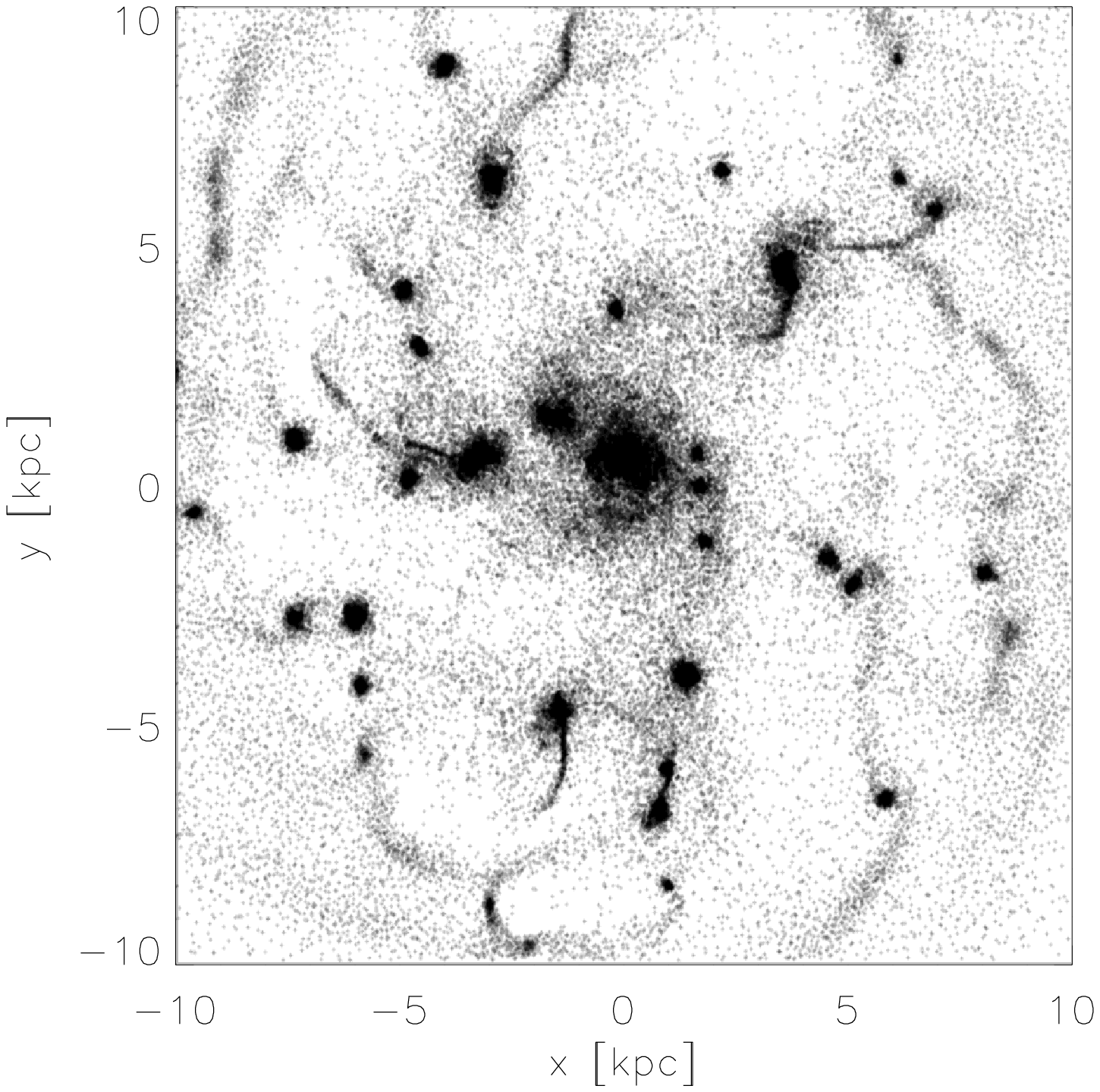}
\includegraphics[width=75mm,height=55mm]{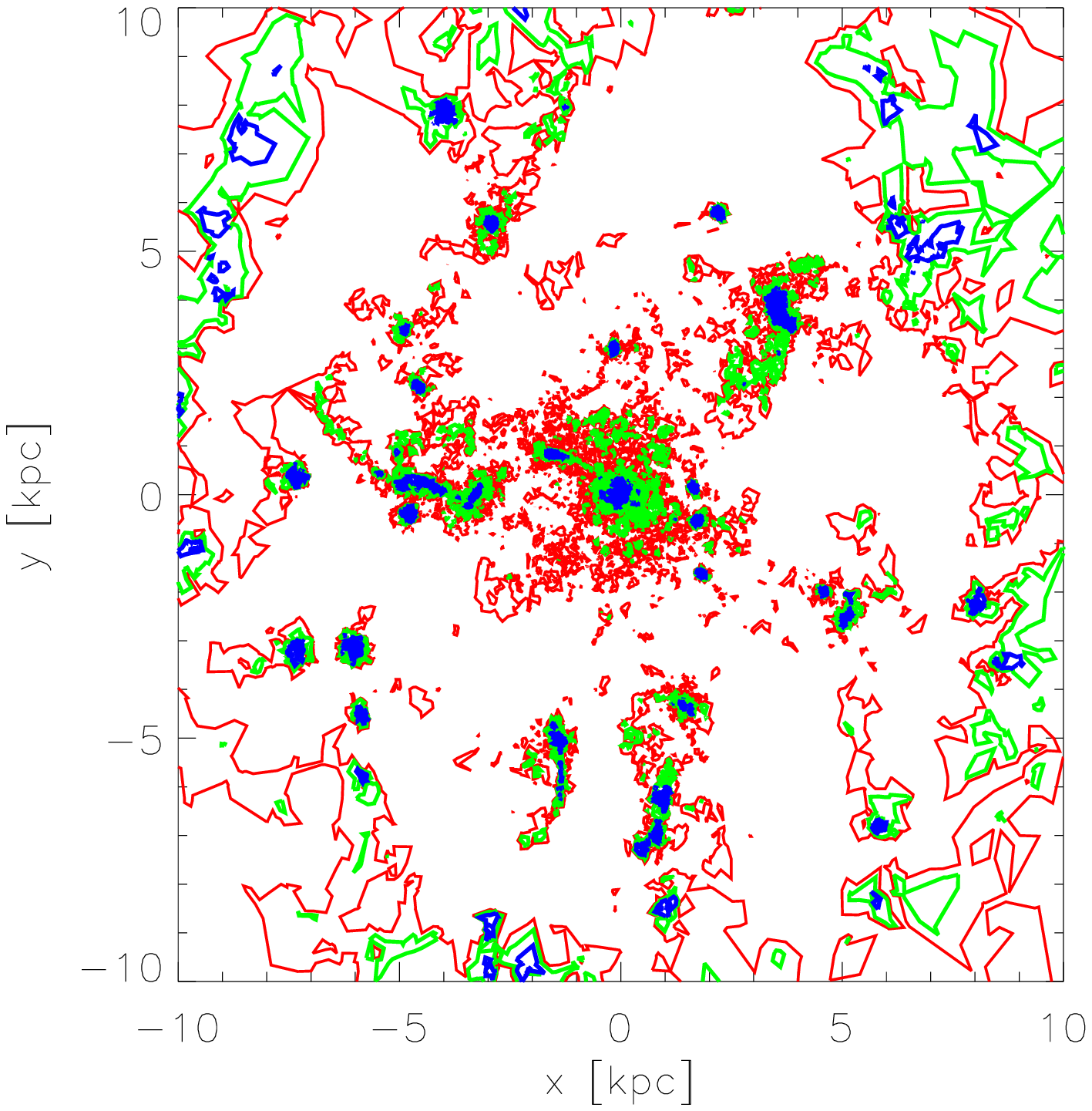}\\
\includegraphics[width=75mm,height=55mm]{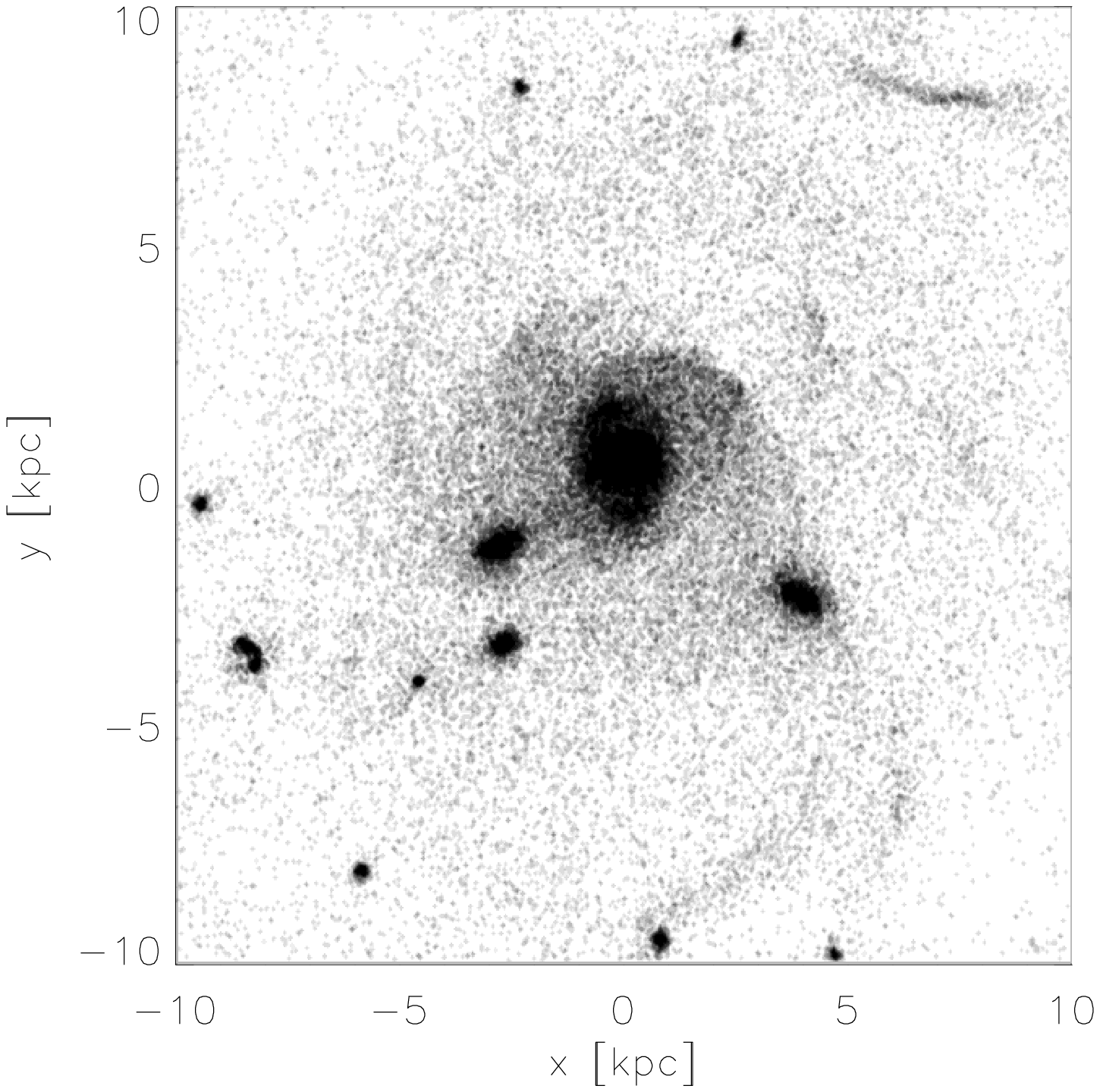}
\includegraphics[width=75mm,height=55mm]{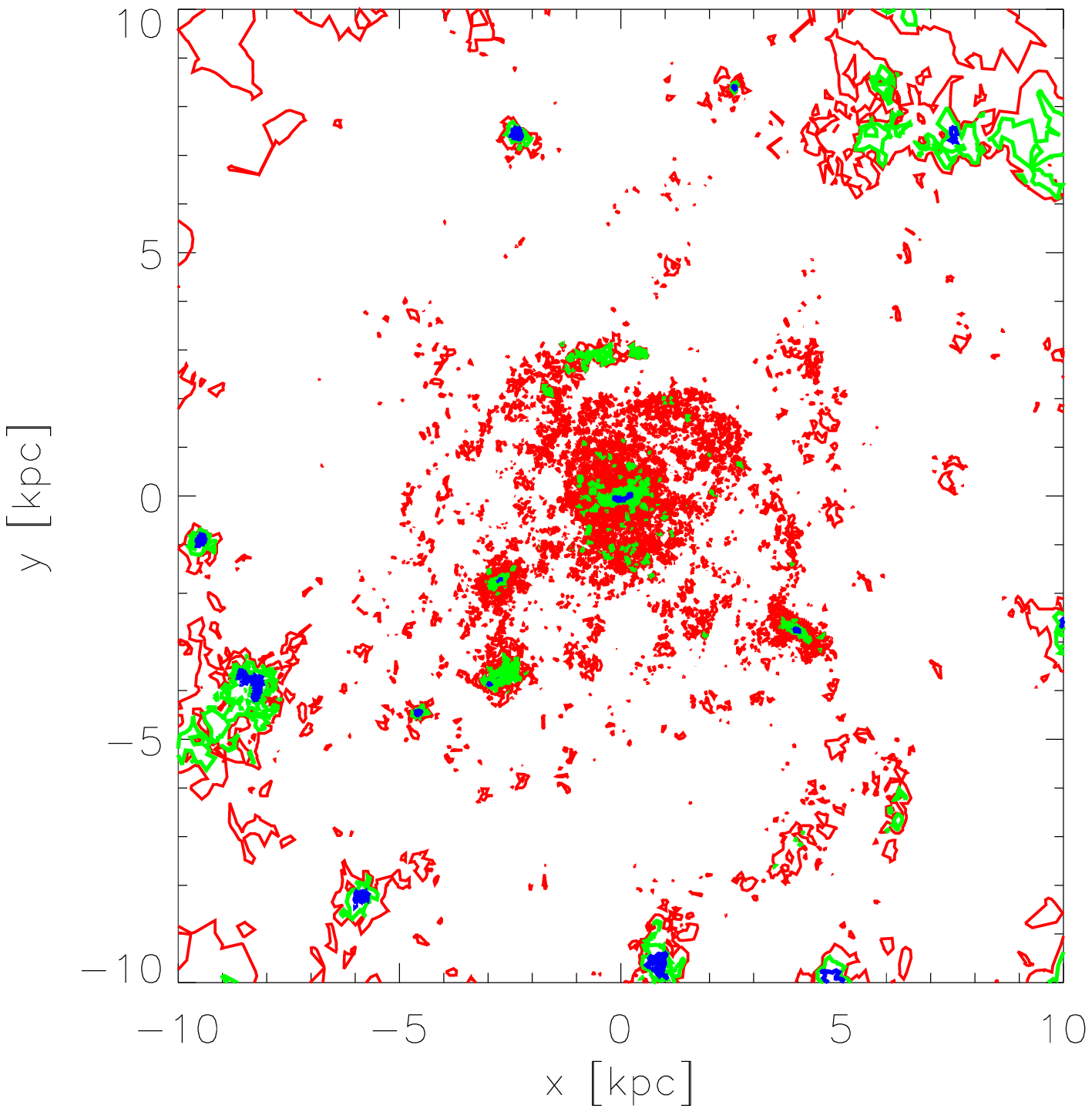}
\caption{
Projected baryonic mass distributions (left-handed panels) and  
colour contours of stellar ages (right-handed panels) for the simulated galaxy disc in S.FeMu experiment at an early and advanced  stages of evolution (lower and upper panels).
  Old, intermediate and young populations are indicated by red, green and blue lines,
  respectively.}
\label{models1}
\end{figure*}

In Fig.~\ref{models1} we show the the projected baryonic mass distributions for S.FeMu at two different
times (after 0.25 and 0.60 Gyr of evolution, upper and lower panels).  We also displayed  
 colour contours defined according to star particle ages.
It is clear from this figure that 
the youngest stellar population are located mainly in clumps. 
We can also see a stellar age gradient 
 which is consistent with a migration scenario 
\citep{guo11}. 
Effectively, by inspecting the time evolution of the galaxy, we find  that clumps migrate 
to the central region of the galaxy on timescales that,
 on average, are about $0.5$ Gyrs,  in agreement with previous analytical and observational results \citep{dekel09, guo11, genzel11}.

In order  to study the origin and survival of clumps in our simulations, 
we explore the ability of different ISM and SN feedback models 
to form and preserve clump structures in disc simulations. 
We find that in the basic SPH model, 
the formation of clumps is prevented because 
gravitationally instability on the gaseous discs is more difficult 
to developed due to the over-smoothing of the density and temperature distributions. 
Conversely, in all our simulations with a multiphase ISM model, 
  gravitational disc instabilities (Toomre 1964) are promoted as a 
  consequence of  a better description of density and temperature 
  gradients \citep{scannapieco06}. 
 The growth of this instability drives the fragmentation of discs in 
clumps which locally match  $Q_{Toomre} < 1$.

 We compare the density and temperature distributions of gas particles 
in our basic SPH simulation (S.BasicSPH) 
with those in the multiphase model (S.Mu). 
The multiphase ISM produces a more 'clumpy' 
and concentrated gas distribution 
 with at least three order of magnitude larger densities 
than the basic SPH.  As expected, these larger  over-densities
 are found to match the clump distribution and are formed by cold gas ($\simeq 10^{4}$ K$^{}$).
   Within these clumpy gas concentrations, 
densities are so large that their cooling times are significantly shorter 
compared to their dynamical time. Consequently, 
gaseous clumps in our multiphase ISM model are transformed  
 in bound stellar systems which, capable to survive the disc shearing, 
migrate and coalesce to the galactic centre. The inclusion of SN feedback 
regulates their stellar masses by heating the gas and triggering outflows as shown below.
Hence, our main analysis will be  focused on  simulations with the multiphase ISM model 
 (S.Mu, S.FeMu, S.ModFeMu, S.StrFeMu), 
unless specifically stated. 

\begin{figure}
\includegraphics[width=86mm,height=72mm]{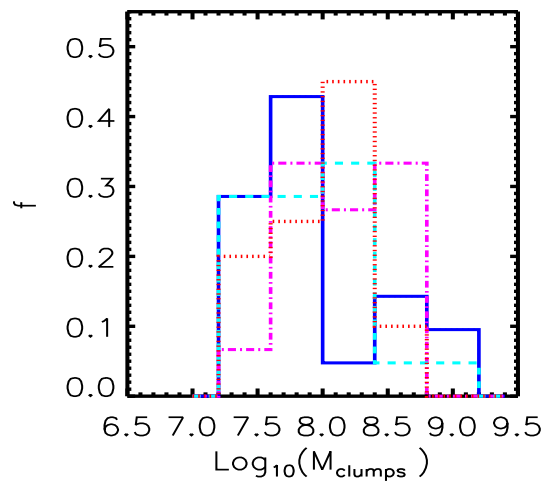}\\
\caption{
   Histograms of stellar masses for clumps in  S.FeMu (blue, solid line),
    S.Mu (magenta, dotted-dashed line), S.ModFeMu (cyan, dashed line) and S.StrFeMu (red, dotted line).
   }
\label{models2}
\end{figure}

Fig.~\ref{models2} shows the  clump mass distributions for experiments  
with the multiphase ISM model and  different SN feedback energy parameters: 
without SN feedback (magenta), with our mild SN feedback model ($E_{SN}= 0.5e51$),  
 with a moderate  ($E_{SN}= 0.7e51$) and a strong ($E_{SN}= 1e51$) SN feedback at a reference time as an example.  
 We find that clumps can survive SN winds even 
in the simulation with strong SN feedback.  
 However, their growths are limited by the SN energy release: the highest mass clumps are found in
the simulation without SN feedback and in the mild SN feedback run. The increase of  the SN energy event produces the
decrease in the average clump mass as expected as the gas is heated up and partially blown away. 
Note also that clumps are continuously accreting new material along their evolutionary path; they are not close systems
but they highly interact with their surrounding ISM.
Stronger SN outflows also contribute to the formation of less gravitational bounded clumps which
can be more easily disrupted.  We find that 
  the mean bound energy for the strong SN feedback run is $\sim 58 $ per cent of that 
corresponding to our mild SN feedback model.
 It is  important to note that the SN feedback models used in this work 
are capable to reproduce the observed 
galactic mass-loading factors \citep{scannapieco06} as well as  
 those of individual clumps  such that the  
mass-loss rates typically exceed the star formation rates by 
a factor of a few \citep{genzel11}. For clumps in the strong SN feedback model, 
we report the highest extreme mass-loading factors of $\sim 7$,  which nevertheless are   
within observed values \citep{genzel11}.

 Clumps in our simulations 
are not fully disrupted by the action of SN feedback as reported by other authors. 
This discrepancy could stem from differences in  the numerical codes, 
which  might lie on the  details in the ISM and feedback models. 
Firstly, we  note that our code does not include other sorts of
feedback apart from thermal heating from SNe. Radiation pressure has been claimed to be 
the dominant mechanism over other sources of feedback, including the SN thermal heating \citep{dekel09, murray10}. Numerical results from  
 \citet{hopkins11} and  \citet{genel12} support this claim. 
  However, its role  in disrupting clumps  remains debated 
\citep{krumholz10, Krumholz2012}. 
 Since we have not a physically-motivated radiative pressure so far implemented in our code, we follow \cite{bournaud11},  and explore a 
stronger SN feedback model (S.StrFeMu) as a way to somehow mimic an extra energy contribution 
for other possible sources,  finding no significant variations in our conclusions.

The SN feedback model developed by  
\citet{scannapieco05, scannapieco06} is one of the most
physically-motivated currently available. In this model,  the SN energy is fractionally distributed into
the gaseous neighbours of two different ISM phases, denoted as {\it hot} and {\it cold}. 
The energy injected into the hot phase is instantaneously 
thermalised and, that received by the cold phase is stored in a {\it reservoir},
 defined for each gas particle in this phase. This energy is  accumulated until 
it is enough to modify the thermodynamical properties of these
particles so they match the corresponding ones of the hot phase . When this happens, the cold particle is
promoted to the hot phase dumping its reservoir energy into its internal energy. 
This scheme prevents artificial losses of SN energy by the cold phase and ensures 
the triggering of mass-loaded galactic winds  which are capable to reproduce the 
observations \citep{genzel11}.

Beyond differences in the feedback schemes, our simulations distinguish
themselves by the adopted multiphase scheme  \citep{scannapieco06}  
as discussed before. 
In S.Mu experiments which includes the
multiphase ISM but not the SN feedback, 
clumps consume their gas into stars, 
 persisting ($\sim$ 0.5 Gyr) as a bound stellar system until coalesce in
the galactic center.  
  The inclusion of SN feedback in our models (S.FeMu, S.ModFeMu and S.StrFeMu)  
partially limits their mass growth and weakly reduces  their lifetimes, 
even though is able to reproduce the observed mass-loading factor of clumps.  
The survival of clumps for these experiments could be explained 
because the SN feedback in our model 
self-regulates the conversion of gas into stars 
in such a way that the dimensionless  star-formation rate efficiency, 
$e_{ff}$, is found to be within the survival regime \citep{krumholz10}. 
In brief, all our simulations with multiphase ISM model promotes  
disc instabilities driving the formation of clumps which migrate and shape  
 a classical bulge.

\subsection{The structure of the bulges}
In order to quantify the role of SN feedback  
in the bulge formation via clumps, we analyse the final structure of bulges 
for S.BasicSPH, S.Mu, S.FeMu, S.ModFeMu and  S.StrFeMu discs, by computing
their  stellar density profiles  and  performing a disc-bulge decomposition by using 
an exponential profile for the disc and a Sersic law for the bulge.
The decomposition was carried out in  radial range internally limited by the gravitational softening and extended up to 8 kpc. 
Note that we use the relation: $bn = 2\times n+0.32$  (Mac Arthur et al 2003) 
 reducing the degree of Sersic parameter space in order to make the fitting process more efficient. 

\begin{figure}
\centering
\includegraphics[width=86mm,height=72mm]{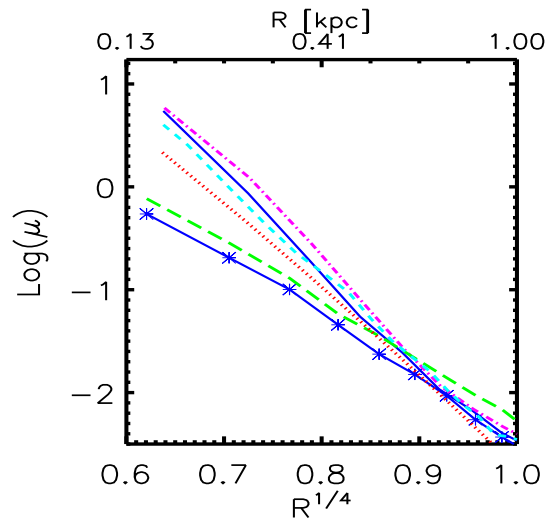}
\caption{Stellar surface density profiles in the central region of simulated galaxies in
 S.BasicSPH (green, long dashed line), S.FeMu (blue, solid line),
    S.Mu (magenta,  dotted-dashed line), S.ModFeMu (cyan, dashed line) and S.StrFeMu (red, dotted line).
    The initial stellar density profile  is plotted
    for comparison (asterisks and blue solid line).  See Table 1 for initial and final Sersic indexes.}
\label{models3}
\end{figure}

Fig.~\ref{models3} shows these stellar density profiles  plotted as a function  r$^{1/4}$ with the aim at 
emphasizing  deviations from classical bulges, principally in 
the central regions \citep{fisher08}.   
According to this figure and the derived Sersic indexes (Table~\ref{tab1}), 
we find that  pseudo-bulges ({\it n} $<$ 2) are only formed when using 
the basic SPH technique (with no multiphase ISM or SN feedback model). 
Classical-bulges ({\it n} $>$ 2) are always formed when  our  multiphase ISM model is switched on.
We analysed the gas and temperature distributions of the run  without and with the multiphase ISM model, finding that this is
capable of reproducing much better density and temperature gradients allowing the co-existence of clouds with different
entropies, generating a turbulence medium which better represents the underlying physics \citep{scannapieco06}.

When the SN feedback model is switched on the growth of the clumps are regulated by the SN energy released by 
event as already shown in Fig.~\ref{models2}.  As a consequence, when the SN energy increased, the bulge profiles get flatter, but even in the strong SN feedback run the resulting bulge has {\it n} $>$ 2. 
We also find that the ratio $B/T$ decreases significantly in the case of the strong SN feedback model as the outflows are more 
violent and  can remove gas or  even
prevent new gas infall more easily.
 These results suggest that our  SN feedback model can regulate the  
 growth of clumps but it does not preclude the building of a classical bulge if sensitive SN energy values are adopted.  
Hence, classical bulges could emerge in a secular scenario by 
clump migration and not only from mergers.

\section{Multi-channel Bulge Formation} 

\begin{figure}
\centering
\includegraphics[width=85mm,height=80mm]{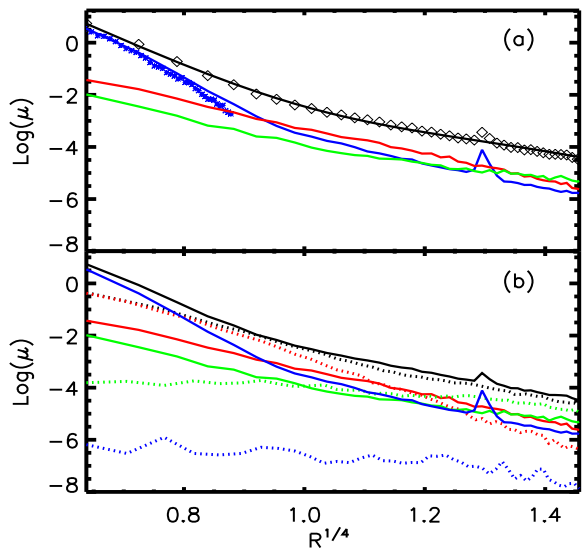}
\caption{
  (a) Final density profiles of the simulated disc for the S.FeMu experiment. 
  Total stellar component (i.e.,  old bulge, old disc and  new stellar 
  population) produces a profile shown by black diamonds, 
  fitted by a  n$\sim$4 Sersic-Exponential function (solid black line).
  Contributions from old bulge, old disc  and  from the new stellar population 
  were discriminated (red, green and blue lines, respectively).   
  Blue asterisks show new stars identified at the final stage  within the 
  bulge but formed in any of the clumps along the galaxy evolution.  
  (b) Time evolution of  density profiles for each stellar components  
  in S.FeMu: Total stellar components (black), old disc (green), old bulge (red) 
  and  new stars (blue). Solid (dotted) lines show the final 
  (initial) profiles.  Thus, dotted blue line represents 
  the new stars formed during the initial snapshot. 
}
\label{SF}
\end{figure}

As mentioned in the previous section,  clump migration should be considered as an alternative
channel of bulge formation. In order to disentangle its contribution, 
we investigate how clumps are able to modify the final density profiles of our S.FeMu disc. 
Fig.~\ref{SF} shows how the different stellar components 
contribute to the final  density profile of the galaxy. 
Note that the stellar population initially distributed 
in the bulge component will be hereinafter refereed as old bulge, 
 and, analogously, for the old disc. These old stars can be  
followed along the galaxy evolution and thus, separated from 
the younger stellar population formed out from the gaseous disc and located 
in the emerging new bulge and disc components. 

 Fig.~\ref{SF}a shows that the new stellar populations (solid blue line) 
significantly contribute to the final density profile in the central region,  
 but also indicates that the new stars formed in clumps (blue asterisk)
are those which have the major role in shaping the bulge.  Note, however,  
 that we specifically excludes the new stars formed in any clump eventually developed
  in situ in the bulge, i.e. we only consider the new stars formed in clumps identified 
on the disc\footnote{ Note that  we use a morphological  bulge-disc decomposition, 
 defining the radius of bulge as twice the effective radius as obtained from the 
Sersic-exponential fitting of profiles. (blue asterisk)
.}

In agreement with previous works  \citep[e.g.,][]{elmegreen08, ceverino10}, 
our results indicate that a secular process of bulge formation, i.e. the clump migration,  
allows the formation of  a classical-like  bulge with a Sersic index of  $n \sim 3-4$  
(see Table~\ref{tab1}). Note that this result seems to be in tension those of 
\cite{bournaud11}, where a central bulge
but with less steeper profile, $n \sim 1.7$, is reported. 
This discrepancy likely emerges from the fact that 
their kinetic feedback might be more efficient destroying clumps  
which, on the other hand, is consistent with the relatively young stellar 
population of clumps found in the latest phases of their merging experiments. 
Also note, that they find the Sersic index by fitting the profile 
from 1.6 kpc ($0.3*R_{1/2}$), much further out than our fitting range.

We also study the relative contribution 
of the new stars and a preexisting bulge and a disk component to the final bulge structure,
including the time evolution of the different stellar density profiles 
in S.FeMu (Fig.~\ref{SF}b). The result indicates that stars distributed initially 
in the Old Bulge gradually become  less dominant, while  new stars 
(formed primarily in clumps) have a more prominent role. 
 This suggests a dynamical interaction between both stellar populations,
which is sculptured in their final relative mass distribution.

  Besides clump migration, other  processes as
bars and mergers are well known to shape the bulge mass distribution. 
Several numerical works have contributed to show that 
mergers are able to form classical bulges ({\it n} $\sim$ 3) and bars induce 
pseudo-bulge ({\it n} $\leq$ 2) formation.
 However, the presence of clumps in these processes
have been little explored. 
 Particularly,  our motivation, as a first approach to the subject, 
is to contribute to answer the question: 
if clump migration by itself develops classical bulges,  
does the combination with other mechanisms make even more difficult to explain
 bulge formation  in a $\Lambda$ Cold Dark Matter cosmology?,
i.e. it will produce an even large excess of classical bulges 
in spiral galaxies at low redshift \citep[see ][]{bournaud11}.

\begin{figure}
\centering
\includegraphics[width=70mm,height=55mm]{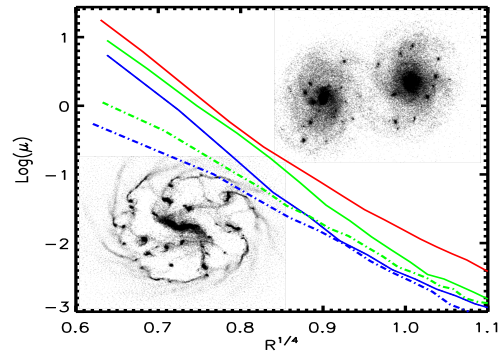}
\caption{ Final stellar density profiles (solid lines):   
S.FeMu (isolated disc, blue), S.FeMu$_{-}$Int (merger remnant, red)  
and S.FeMu$_{-}$Bar (bar, green). Initial profiles are shown for reference (dash-dotted lines). 
Note that for construction, the initial profile of the merger remnant 
coincides with that of the isolated disc.
It is remarkable that models with 
similar Sersic index value like S.FeMu  and S.FeMu$_{-}$Int ($n \sim 3-4$) came out 
of very different dynamical histories.}
\label{IntBar}
\end{figure}

In order to explore this topic, we investigate 
how these mechanisms might compete with each other or reinforce 
the action of clumps by studying their 
contributions to the  density profiles. 
In Fig.~\ref{IntBar}, we show  final profiles of
 the isolated disc of S.FeMu, of its associated model
with bars (S.FeMu$_{-}$Bar) and of the merger remnant
in S.FeMu$_{-}$Int.   The bulge growth  can be followed by comparing 
final density profiles (solid lines) with their respective initial ones 
 (dash-dotted  lines), or 
more quantitative, by  their Sersic indexes 
(see Table~\ref{tab1}). 

Our results indicate that  clump migration in isolated discs can be  
as effective as mergers in developing a classical-bulge (high Sersic index),  
but with a less extended bulge growth than mergers, i.e. lower effective 
radius and bulge-to-total stellar masses (see Table 1).

 Our model with a more dominant  disk (S.FeMu$_{-}$Bar) presents a different evolution 
compared with the 
 S.FeMu. As all the simulated cases, the gas rich-disk 
 fragments generating clumps, but the model also develops 
relatively soon a large scale bar-like structure, 
similar to that observed in some 
 barred galaxies with a clumpy stellar distribution \citep {Hernandez11}. 
The gas distribution and motion is affected by this bar triggering an inflow, 
which  in turn  weakens the large scale bar.
This behaviour is not unusual in gaseous bar simulations \citep {Norman1996, immeli04}.

It is well known that bar robustness is highly sensitive to 
the numerical time step \citep {Shen2004, Klypin2009}. 
In order to check the robustness of this trend,  
we re-run the model S.FeMu$_{-}$Bar with a smaller time step. 
In both  of our experiments a large scale bar forms and weakens. 
 The overall result of our barred experiments is that bar formation produce 
a less compact pseudo-bulge with index $n \sim 2.3$, even when clumps form. 
Our  result,  could be explained as  the  dominant effect of 
the large scale bar which induces a mass redistribution 
over the contribution of clump migration.  
  The pictures emerging out  of our analysis may provide an explanation to  the study recently reported by  \cite{Okamoto2012},  where bar formation and clumpy star formation co-exist in a simulated galaxies within a  cosmological context,  with a final bulge with a low Sersic index and a dominant mass contribution 
from a central starburst   \citep[also see,][]{Shigeki2011}. It is important to say that a systematic scan of the parameter space is required in order to accurately characterize this evolution channel.

\section{Conclusions}

We used hydrodynamical simulations of isolated gas-rich discs crafted to mimic 
star forming galaxies at z$\sim$2-3. The simulations include a realistic SN feedback model 
implemented with a multiphase treatment  of the ISM \citep{scannapieco05, scannapieco06},  
 which has been proven to be able to describe galactic global properties comparable
to observations \citep{Scannapieco2008, Scannapieco2009, derossi2011}.

 We find that the presence of the inhomogeneous, turbulent multiphase medium in our simulations 
promotes the formation and growth of  gravitational disc instability 
favouring  stellar clumps formation. 
For our specific multiphase ISM and SN feedback model, 
the gas in clumps is transformed into stars 
 at a rate of a less than one per cent of the clump mass per free-fall time, 
producing bound systems within the survival regime \citep{krumholz10},
and stellar age spread of the order of 500 Myr. This indicates that 
 the survival of clumps is not strongly affected by our
SN feedback model, if sensitive values are adopted for the SN energy released.
  Our SN feedback model is also able to reproduce the observed 
 mass-loading factors of clumps  \citep{genzel11}.

The picture which emerges out from the analysis of our numerical experiments is that bulge formation 
by clumps may co-exist with other channels of bulge assembly  producing 
realistic values of Sersic index  and bulge-to-total mass ratios. 
 As a consequence, clump coalescence becomes  a viable channel for bulge formation.
In this scenario, bulges  could be interpreted as   compound systems similar to those observed in the Milky Way \citep{Babusiaux} 
and some external galaxies \citep{fisher08},  with different relative contributions 
 of pre-existing bulge, old disc and new stars, which may store archaeological information 
of the galaxy assembling.

\section*{Acknowledgments}
We thanks the anonymous referee for useful comments which help to improve this paper.
This work was partially supported by 
 Consejo Nacional de Investigaciones Cient\'{\i}ficas
y T\'ecnicas (PIP 2009/0305),  PICT Raices 2011 (959) and Programa de Cooperaci\'on
Cient\'{\i}fico-Tecnol\'ogica MINCyT-CONACyT (MX/09/09). 
The simulations are part of the
SouthCross project and  were run in HOPE and Fenix clusters at IAFE, Argentina.



\bibliographystyle{mn2e}

\def\apj{ApJ}
\def\apjl{ApJ}
\def\aj{AJ}
\def\mnras{MNRAS}
\def\aa{A\&A}
\def\nat{nat}
\def\araa{ARA\&A}
\def\aap{A\&A}


\bibliography{clumpsFinal3}

\begin{thebibliography}{}

\bibitem[\protect\citeauthoryear{{Babusiaux}, {G{\'o}mez}, {Hill}, {Royer},
  {Zoccali}, {Arenou}, {Fux}, {Lecureur}, {Schultheis}, {Barbuy}, {Minniti} \&
  {Ortolani}}{{Babusiaux} et~al.}{2010}]{Babusiaux}
{Babusiaux} C.,  {G{\'o}mez} A.,  {Hill} V.,  {Royer} F.,  {Zoccali} M.,
  {Arenou} F.,  {Fux} R.,  {Lecureur} A.,  {Schultheis} M.,  {Barbuy} B.,
  {Minniti} D.,    {Ortolani} S.,  2010, AAP, 519, 77

\bibitem[\protect\citeauthoryear{{Bournaud}, {Chapon}, {Teyssier}, {Powell},
  {Elmegreen}, {Elmegreen}, {Duc}, {Contini}, {Epinat} \& {Shapiro}}{{Bournaud}
  et~al.}{2011}]{bournaud11}
{Bournaud} F.,  {Chapon} D.,  {Teyssier} R.,  {Powell} L.~C.,  {Elmegreen}
  B.~G.,  {Elmegreen} D.~M.,  {Duc} P.-A.,  {Contini} T.,  {Epinat} B.,
  {Shapiro} K.~L.,  2011, \apj, 730, 4

\bibitem[\protect\citeauthoryear{{Bournaud}, {Elmegreen} \&
  {Elmegreen}}{{Bournaud} et~al.}{2007}]{bournaud07}
{Bournaud} F.,  {Elmegreen} B.~G.,    {Elmegreen} D.~M.,  2007, \apj, 670, 237

\bibitem[\protect\citeauthoryear{{Ceverino}, {Dekel} \& {Bournaud}}{{Ceverino}
  et~al.}{2010}]{ceverino10}
{Ceverino} D.,  {Dekel} A.,    {Bournaud} F.,  2010, \mnras, 404, 2151

\bibitem[\protect\citeauthoryear{{De Rossi}, {Tissera} \& {Pedrosa}}{{De Rossi}
  et~al.}{2011}]{derossi2011}
{De Rossi} M.~E.,  {Tissera} P.~B.,    {Pedrosa} S.~E.,  2011, Boletin de la
  Asociacion Argentina de Astronomia La Plata Argentina, 54, 373

\bibitem[\protect\citeauthoryear{{Dekel}, {Sari} \& {Ceverino}}{{Dekel}
  et~al.}{2009}]{dekel09}
{Dekel} A.,  {Sari} R.,    {Ceverino} D.,  2009, \apj, 703, 785

\bibitem[\protect\citeauthoryear{{Elmegreen}, {Bournaud} \&
  {Elmegreen}}{{Elmegreen} et~al.}{2008}]{elmegreen08}
{Elmegreen} B.~G.,  {Bournaud} F.,    {Elmegreen} D.~M.,  2008, \apj, 688, 67

\bibitem[\protect\citeauthoryear{{Fisher} \& {Drory}}{{Fisher} \&
  {Drory}}{2008}]{fisher08}
{Fisher} D.~B.,  {Drory} N.,  2008, \apj, 136, 773

\bibitem[\protect\citeauthoryear{{F{\"o}rster Schreiber}, {Genzel},
  {Bouch{\'e}}, {Cresci}, {Davies}, {Buschkamp}, {Shapiro}, {Tacconi}, {Hicks},
  {Genel} \& {Shapley}}{{F{\"o}rster Schreiber} et~al.}{2009}]{forster09}
{F{\"o}rster Schreiber} N.~M.,  {Genzel} R.,  {Bouch{\'e}} N.,  {Cresci} G.,
  {Davies} R.,  {Buschkamp} P.,  {Shapiro} K.,  {Tacconi} L.~J.,  {Hicks}
  E.~K.~S.,  {Genel} S.,    {Shapley} A.~E.,  2009, \apj, 706, 1364

\bibitem[\protect\citeauthoryear{{F{\"o}rster Schreiber}, {Shapley}, {Genzel},
  {Bouch{\'e}}, {Cresci}, {Davies}, {Erb}, {Genel}, {Lutz}, {Newman},
  {Shapiro}, {Steidel}, {Sternberg} \& {Tacconi}}{{F{\"o}rster Schreiber}
  et~al.}{2011}]{Forster2011}
{F{\"o}rster Schreiber} N.~M.,  {Shapley} A.~E.,  {Genzel} R.,  {Bouch{\'e}}
  N.,  {Cresci} G.,  {Davies} R.,  {Erb} D.~K.,  {Genel} S.,  {Lutz} D.,
  {Newman} S.,  {Shapiro} K.~L.,  {Steidel} C.~C.,  {Sternberg} A.,
  {Tacconi} L.~J.,  2011, \apj, 739, 45

\bibitem[\protect\citeauthoryear{{Genel}, {Naab}, {Genzel}, {F{\"o}rster
  Schreiber}, {Sternberg}, {Oser}, {Johansson}, {Dav{\'e}}, {Oppenheimer} \&
  {Burkert}}{{Genel} et~al.}{2012}]{genel12}
{Genel} S.,  {Naab} T.,  {Genzel} R.,  {F{\"o}rster Schreiber} N.~M.,
  {Sternberg} A.,  {Oser} L.,  {Johansson} P.~H.,  {Dav{\'e}} R.,
  {Oppenheimer} B.~D.,    {Burkert} A.,  2012, \apj, 745, 11

\bibitem[\protect\citeauthoryear{{Genzel}, {Burkert}, {Bouch{\'e}}, {Cresci},
  {F{\"o}rster Schreiber}, {Shapley}, {Shapiro}, {Tacconi}, {Buschkamp},
  {Cimatti}, {Daddi}, {Davies}, {Eisenhauer}, {Erb} \& {Genel}}{{Genzel}
  et~al.}{2008}]{genzel08}
{Genzel} R.,  {Burkert} A.,  {Bouch{\'e}} N.,  {Cresci} G.,  {F{\"o}rster
  Schreiber} N.~M.,  {Shapley} A.,  {Shapiro} K.,  {Tacconi} L.~J.,
  {Buschkamp} P.,  {Cimatti} A.,  {Daddi} E.,  {Davies} R.,  {Eisenhauer} F.,
  {Erb} D.~K.,    {Genel} S.,  2008, \apj, 687, 59

\bibitem[\protect\citeauthoryear{{Genzel}, {Newman}, {Jones}, {F{\"o}rster
  Schreiber}, {Shapiro}, {Genel}, {Lilly}, {Renzini}, {Tacconi}, {Bouch{\'e}},
  {Burkert}, {Cresci}, {Buschkamp}, {Carollo} \& {Ceverino}}{{Genzel}
  et~al.}{2011}]{genzel11}
{Genzel} R.,  {Newman} S.,  {Jones} T.,  {F{\"o}rster Schreiber} N.~M.,
  {Shapiro} K.,  {Genel} S.,  {Lilly} S.~J.,  {Renzini} A.,  {Tacconi} L.~J.,
  {Bouch{\'e}} N.,  {Burkert} A.,  {Cresci} G.,  {Buschkamp} P.,  {Carollo}
  C.~M.,    {Ceverino} D.,  2011, \apj, 733, 101

\bibitem[\protect\citeauthoryear{{Guo}, {Giavalisco}, {Ferguson}, {Cassata} \&
  {Koekemoer}}{{Guo} et~al.}{2011}]{guo11}
{Guo} Y.,  {Giavalisco} M.,  {Ferguson} H.~C.,  {Cassata} P.,    {Koekemoer}
  A.~M.,  2011, ArXiv e-prints

\bibitem[\protect\citeauthoryear{{Hern{\'a}ndez-Toledo}, {Cano-D{\'{\i}}az},
  {Valenzuela}, {Puerari}, {Garc{\'{\i}}a-Barreto}, {Moreno-D{\'{\i}}az} \&
  {Bravo-Alfaro}}{{Hern{\'a}ndez-Toledo} et~al.}{2011}]{Hernandez11}
{Hern{\'a}ndez-Toledo} H.~M.,  {Cano-D{\'{\i}}az} M.,  {Valenzuela} O.,
  {Puerari} I.,  {Garc{\'{\i}}a-Barreto} J.~A.,  {Moreno-D{\'{\i}}az} E.,
  {Bravo-Alfaro} H.,  2011, \aj, 142, 182

\bibitem[\protect\citeauthoryear{{Hopkins}, {Keres}, {Murray}, {Quataert} \&
  {Hernquist}}{{Hopkins} et~al.}{2011}]{hopkins11}
{Hopkins} P.~F.,  {Keres} D.,  {Murray} N.,  {Quataert} E.,    {Hernquist} L.,
  2011, ArXiv e-prints

\bibitem[\protect\citeauthoryear{{Immeli}, {Samland}, {Gerhard} \&
  {Westera}}{{Immeli} et~al.}{2004}]{immeli04}
{Immeli} A.,  {Samland} M.,  {Gerhard} O.,    {Westera} P.,  2004, \aap, 413,
  547

\bibitem[\protect\citeauthoryear{{Klypin}, {Valenzuela}, {Col{\'{\i}}n} \&
  {Quinn}}{{Klypin} et~al.}{2009}]{Klypin2009}
{Klypin} A.,  {Valenzuela} O.,  {Col{\'{\i}}n} P.,    {Quinn} T.,  2009,
  \mnras, 398, 1027

\bibitem[\protect\citeauthoryear{{Krumholz} \& {Dekel}}{{Krumholz} \&
  {Dekel}}{2010}]{krumholz10}
{Krumholz} M.~R.,  {Dekel} A.,  2010, \mnras, 406, 112

\bibitem[\protect\citeauthoryear{{Krumholz} \& {Thompson}}{{Krumholz} \&
  {Thompson}}{2012}]{Krumholz2012}
{Krumholz} M.~R.,  {Thompson} T.~A.,  2012, ArXiv e-prints

\bibitem[\protect\citeauthoryear{{Murray}, {Quataert} \& {Thompson}}{{Murray}
  et~al.}{2010}]{murray10}
{Murray} N.,  {Quataert} E.,    {Thompson} T.~A.,  2010, \apj, 709, 191

\bibitem[\protect\citeauthoryear{{Norman}, {Sellwood} \& {Hasan}}{{Norman}
  et~al.}{1996}]{Norman1996}
{Norman} C.~A.,  {Sellwood} J.~A.,    {Hasan} H.,  1996, \apj, 462, 114

\bibitem[\protect\citeauthoryear{{Okamoto}}{{Okamoto}}{2012}]{Okamoto2012}
{Okamoto} T.,  2012, astroph

\bibitem[\protect\citeauthoryear{{Perez}, {Michel-Dansac} \& {Tissera}}{{Perez}
  et~al.}{2011}]{perez11}
{Perez} J.,  {Michel-Dansac} L.,    {Tissera} P.~B.,  2011, \mnras, 417, 580

\bibitem[\protect\citeauthoryear{{Scannapieco}, {Tissera}, {White} \&
  {Springel}}{{Scannapieco} et~al.}{2005}]{scannapieco05}
{Scannapieco} C.,  {Tissera} P.~B.,  {White} S.~D.~M.,    {Springel} V.,  2005,
  \mnras, 364, 552

\bibitem[\protect\citeauthoryear{{Scannapieco}, {Tissera}, {White} \&
  {Springel}}{{Scannapieco} et~al.}{2006}]{scannapieco06}
{Scannapieco} C.,  {Tissera} P.~B.,  {White} S.~D.~M.,    {Springel} V.,  2006,
  \mnras, 371, 1125

\bibitem[\protect\citeauthoryear{{Scannapieco}, {Tissera}, {White} \&
  {Springel}}{{Scannapieco} et~al.}{2008}]{Scannapieco2008}
{Scannapieco} C.,  {Tissera} P.~B.,  {White} S.~D.~M.,    {Springel} V.,  2008,
  \mnras, 389, 1137

\bibitem[\protect\citeauthoryear{{Scannapieco}, {White}, {Springel} \&
  {Tissera}}{{Scannapieco} et~al.}{2009}]{Scannapieco2009}
{Scannapieco} C.,  {White} S.~D.~M.,  {Springel} V.,    {Tissera} P.~B.,  2009,
  \mnras, 396, 696

\bibitem[\protect\citeauthoryear{{Shen} \& {Sellwood}}{{Shen} \&
  {Sellwood}}{2004}]{Shen2004}
{Shen} J.,  {Sellwood} J.~A.,  2004, \apj, 604, 614

\bibitem[\protect\citeauthoryear{{Shigeki} \& {Takayuki}}{{Shigeki} \&
  {Takayuki}}{2011}]{Shigeki2011}
{Shigeki} I.,  {Takayuki} R.~S.,  2011, ArXiv e-prints

\end{thebibliography}
\IfFileExists{\jobname.bbl}{}
{\typeout{}
\typeout{****************************************************}
\typeout{****************************************************}
\typeout{** Please run "bibtex \jobname" to optain}
\typeout{** the bibliography and then re-run LaTeX}
\typeout{** twice to fix the references!}
\typeout{****************************************************}
\typeout{****************************************************}
\typeout{}
}

\bsp

\label{lastpage}

\end{document}